# Continuous variables triple-photon states quantum entanglement

E. A. Rojas González,[1,2] A. Borne,[3,4] B. Boulanger,[4] J. A. Levenson,[1] and K. Bencheikh[1,*]

[1]*Centre de Nanosciences et de Nanotechnologies CNRS/Université Paris-Saclay, 91460 Marcoussis, France*
[2]*Department of Engineering Sciences, Uppsala University, Uppsala, Sweden*
[3]*Weizmann Institutes of Science, Rehovot 7610001, Israel*
[4]*Institut Néel, CNRS/Université Grenoble Alpes, 38402 Grenoble, France*
(Dated: September 11, 2017)

We investigate the quantum entanglement of the three modes associated with the three-photon states obtained by triple-photon generation in a phase-matched third-order nonlinear optical interaction. Although the second order processes have been extensively dealt with, there is no direct analogy between the second and third-order mechanisms. We show for example the absence of quantum entanglement between the quadratures of the three modes in the case of spontaneous parametric triple-photon generation. However, we show that genuine triple-photon entanglement is obtained in the fully seeded case, and its efficiency increases with the seeding level.

Twin-photon states produced by second-order nonlinear optical parametric interactions have deeply influenced the history of quantum optics and are the main ingredient in several quantum protocols. Their ubiquity contrasts with the case of triple-photon states (TPS). The unique quantum properties of TPS were anticipated in the 1989 seminal paper of Greenberger, Horne and Zeilinger, from which a particular category of TPS is called GHZ [1]. Their tripartite entanglement is identified as a direct way to contradict local realism theory for non-statistical predictions of quantum mechanics. Later on, W states have been predicted to retain maximally bipartite entanglement when any one of the three qubits is traced out [2]. Very recently, anisotropy was identified as a new invariant for pure three-qubit states and was found identical for each of their pairs, opening several promising applications [3]. More generally, one can expect that TPS plays in quantum optics a similar role as twin-photon states did over the last 40 years and allow the development of novel and efficient quantum protocols based on triple-photon entanglement or announced entangled twins. This is provided that clever optimization of both material and nonlinear interaction is achieved to overcome the well-known weakness of third-order nonlinear processes. Until now, only configurations based on second-order nonlinear interactions have succeed in producing TPS [4–8] at the expense of increased complexity, low efficiency or detection conditioned protocols preventing their use. However, the most direct way to produce TPS is indeed in materials exhibiting a pure third-order nonlinear susceptibility, $\chi^{(3)}$, in which a pump photon with energy $\hbar\omega_p$ is down-converted simultaneously into a triplet $\hbar\omega_1$, $\hbar\omega_2$ and $\hbar\omega_3$ with the subsequent energy conservation, $\hbar\omega_p = \hbar\omega_1 + \hbar\omega_2 + \hbar\omega_3$, and momentum conservation $\hbar\vec{k}_p = \hbar\vec{k}_1 + \hbar\vec{k}_2 + \hbar\vec{k}_3$. Only one experimental demonstration of Triple-Photon Generation (TPG) was reported in the regime of bi-state seeding, i.e., two over the three fields associated to the out-coming component of the triple-photons are injected together with the pump field [9]. This pioneering work at the classical level clearly demonstrates a phase-matched third-order nonlinear process. Its transposition to spontaneous down-conversion is however precluded by the small $\chi^{(3)}$ value. For instance, in the fluorescence regime, less than a triplet per day is expected [10]. There is an active research for novel materials and geometrical configurations pushing nowadays the efficiency of the reverse nonlinear process, namely third-harmonic generation, from bulk to waveguide operation [11–14]. One can reasonably conjecture this will contribute to increase TPG efficiency as predicted by semiclassical theories [15–17]. However, one can wonder if our comprehension at the quantum level of the process is sufficient to optimize the interaction. Until now, this optimization is strongly driven by the analogy with the well-known and largely investigated $\chi^{(2)}$ process.

In this Letter, we develop a full quantum description of TPS in the framework of Continuous Variables (CV) corresponding to the manipulation of the electromagnetic fields. To the best of our knowledge, the CV quantum properties [18] remain unexplored for TPS generated by a pure third-order nonlinear interaction. We discuss spontaneous, single, double, and triple-seeded configurations and demonstrate that using twin-photon knowledge is often a wrong strategy for TPS optimization. As a result, no entanglement is found in the spontaneous emission regime. More importantly, tripartite entanglement is found in the triple seeding configuration and is found strongly increasing with the seeding, becoming a novel avenue for TPS optimization.

The starting point of our theoretical analysis is the interaction Hamiltonian describing TPG in a nonlinear material exhibiting a third-order nonlinear susceptibility [19]. It reduces to

$$\hat{H} = \hbar\kappa(\hat{a}_1^\dagger \hat{a}_2^\dagger \hat{a}_3^\dagger + \hat{a}_1 \hat{a}_2 \hat{a}_3), \qquad (1)$$

for a monochromatic strong non-depleted classical pump.

$\kappa$ is proportional to the pump amplitude and to the nonlinear susceptibility $\chi^{(3)}$. The annihilation operators $\hat{a}_k$ ($k = 1, 2, 3$) describe the triple photons modes. The Hamiltonian describes the annihilation of a pump photon and the simultaneous creation of triple photons, and the inverse process. We now consider the evolution of the system in the Heisenberg picture. The operators time evolution is given by:

$$\frac{d\hat{a}_k(t)}{dt} = \frac{\imath}{\hbar}[\hat{H}, \hat{a}_k(t)]. \qquad (2)$$

Let $\hat{A}_k$ ($\hat{A}_k^\dagger$) be the annihilation (creation) operators of mode $k$ after the nonlinear interaction. We define the amplitude and phase quadratures $\hat{p}_k = \hat{A}_k + \hat{A}_k^\dagger$ and $\hat{q}_k = \imath(\hat{A}_k - \hat{A}_k^\dagger)$, satisfying the canonical commutation relations $[\hat{q}_k, \hat{p}_k] = 2\imath$. Their quantum fluctuations are measured with balanced homodyne detections. By sweeping the phase $\theta_k$ of the local oscillator, used to measure mode $k$, one has access to the generalized quadratures $\hat{P}_k(\theta_k) = e^{-\imath\theta_k}\hat{A}_k^\dagger + e^{\imath\theta_k}\hat{A}_k$ and $\hat{Q}_k(\theta_k) = \hat{P}_k(\theta_k + \pi/2)$. Let us finally define the following linear combinations:

$$\begin{aligned}\hat{u} &= h_1\hat{P}_1 + h_2\hat{P}_2 + h_3\hat{P}_3 \\ \hat{v} &= g_1\hat{Q}_1 + g_2\hat{Q}_2 + g_3\hat{Q}_3,\end{aligned} \qquad (3)$$

where $h_k$ and $g_k$ ($k = 1, 2, 3$) are arbitrary real parameters introduced experimentally as electrical attenuations or amplifications of the photocurrents generated by the homodyne detection photodetectors. Phase dependence is intentionally omitted in Eq.(3) for simplicity. We quantify multi-body quantum entanglement of TPS using the non-separability criterion $S$ introduced by P. van Loock and A. Furusawa in Ref.[20], defined as:

$$S = \langle \Delta\hat{u}^2 \rangle + \langle \Delta\hat{v}^2 \rangle, \qquad (4)$$

which can be measured experimentally using homodyne detection technics. Expanding Eq. (4) and using Eq. (3) we find:

$$\begin{aligned}S = &\sum_{k=1}^3 h_k^2 \langle \Delta\hat{Q}_k^2 \rangle + \sum_{k=1}^3 g_k^2 \langle \Delta\hat{P}_k^2 \rangle \\ &+ \sum_{k=1}^3 \sum_{\substack{m=1 \\ m \neq k}}^3 h_k h_m (\langle \hat{Q}_k \hat{Q}_m \rangle - \langle \hat{Q}_k \rangle \langle \hat{Q}_m \rangle) \\ &+ \sum_{k=1}^3 \sum_{\substack{m=1 \\ m \neq k}}^3 g_k g_m (\langle \hat{P}_k \hat{P}_m \rangle - \langle \hat{P}_k \rangle \langle \hat{P}_m \rangle).\end{aligned} \qquad (5)$$

The two first terms are a linear summation of the generalized quadrature variances. The two last terms describe cross-correlations between the output modes at the origin of the entanglement. Indeed, whereas the two first terms are always positive, the two last can be negative and can interfere destructively with the first ones. The criterion $S$ is a function of the real parameters $h_k$ and $g_k$. If

$$S < f_p = 2(|h_k g_k| + |h_l g_l + h_m g_m|) \qquad (6)$$

for a given permutation $\{k, l, m\}$ of $\{1, 2, 3\}$, then the quantum system is fully inseparable and exhibits genuine tripartite quantum entanglement [20]. However, if

$$S < f_s = 2(|h_k g_k| + |h_l g_l| + |h_m g_m|), \qquad (7)$$

which is less restrictive than Eq. (6), the system is at least partially separable. If $f_p \leq S < f_s$, two among the three modes could be entangled, without being entangled to the third part. According to [20], the parameters $h_k$ and $g_k$ are chosen such that $0 < f_p \leq f_s$ and $[\hat{u}, \hat{v}] = 0$, to allow $S \to 0$ and the existence of simultaneous eigenstates of $\hat{u}$ and $\hat{v}$.

The determination of $S$ in the case of TPS relies on the resolution of Eq. (2). Unfortunately, they have no known analytical solutions when solved in the frame of quantum mechanics, though it has been shown that the exact classical solutions are Jacobi elliptic functions [21]. Here, to obtain an approximate solution of the operators at a time $t$, we use the Baker-Hausdorff expansion

$$\hat{A}_k = \hat{a}_k + \sum_{n=1}^\infty \frac{(\imath\xi)^n}{n!} \hat{\Omega}_{n;klm}, \qquad (8)$$

to a finite order, where $\xi = \kappa t$ is the interaction strength and

$$\hat{\Omega}_{n;klm} = \frac{1}{(\hbar\kappa)^n} \underbrace{[\hat{H}, [\hat{H}, [\cdots \hat{H}, [\hat{H}, \hat{a}_k]]]]}_{n\text{-times}}, \qquad (9)$$

with $\{k, l, m\}$ being permutations of $\{1, 2, 3\}$. Expanding the commutators in Eq. (9), we find that $\hat{\Omega}_{n;klm}$ is a function of the creation and annihilation operators of the three modes raised at different powers. We also find a symmetry with respect to the indexes $l$ and $m$, i.e. $\hat{\Omega}_{n;klm} = \hat{\Omega}_{n;kml}$, which is helpful in the subsequent calculations. The expansion order of the operators in Eq. (8) is crucial for the validity of our analysis. It depends on the interaction strength $\xi$ and on the average number of seeding photons. In the following, we will analyze different situations of TPG expanding Eq. (8) to a finite order and calculating the criterion $S$ using a combination of both symbolic and numerical computational methods. For all numerical analysis we take $|\xi| = 1.75 \times 10^{-6}$, deduced from [21], which is very representative of the third-order nonlinearity of nowadays materials.

*Triple-photon parametric fluorescence.* In the fluorescence case, there is no seeding and since $|\xi| \ll 1$, the expansion is valid at any order. It is well known that Spontaneous Parametric Down-Conversion (SPDC) induces strong quantum entanglement between the twin modes. Surprisingly, our analysis shows that a spontaneous triple-photon generation does not exhibit such quantum entanglement in the CV regime for any chosen linear combination of $\hat{u}$ and $\hat{v}$. We find that the cross-correlation terms in Eq. (5) between the output modes vanish and that $S$ reduces to

$$S = (1 + 2\langle \hat{A}^\dagger \hat{A} \rangle) \sum_{k=1}^{3} (h_k^2 + g_k^2), \quad (10)$$

when assuming identical average photon number $\langle \hat{A}^\dagger \hat{A} \rangle$ in each mode. The full derivation of Eq. (10) is given in the supplemental material [22]. According to Eq. (10), the criterion $S$ only depends on the strength of the interaction $\xi$, which is contained in $\langle \hat{A}^\dagger \hat{A} \rangle$ and on the parameters $h_k$ and $g_k$. We compare the result given by Eq. (10) to the classical limit $f_s$ given in Eq. (7) by analyzing the difference $S - f_s$. After some mathematical manipulations, we end up with $S - f_s = 2\Gamma^2 \langle \hat{A}^\dagger \hat{A} \rangle + \sum_{k=1}^{3}(|h_k| - |g_k|)^2$, where $\Gamma^2 = (h_1^2 + h_2^2 + h_3^2 + g_1^2 + g_2^2 + g_3^2)$. Each term being positive, $S \geqslant f_s$ for any choice of $h_k$ and $g_k$. This result clearly demonstrates that the three modes are fully independent, meaning that their quantum fluctuations are totally uncorrelated. Though we were expecting a similarity with respect to the twin-photon fluorescence where an entanglement is found, this result could be easily understood as follows. The triplets are generated from the vacuum quantum fluctuations as the twins. However, for a given eigenvalue outcome of the observable, $\hat{Q}_1$ for example, the two others, $\hat{Q}_2$ and $\hat{Q}_3$, can still take any random pair of eigenvalues. Each realization is thus independent from the previous one.

*Partially seeded triple-photon generation.* The validity of our analysis in the seeded cases is discussed in details in the supplemental material [22], treating particularly the fully seeded TPG which is the most restrictive case. We show that a sufficient condition for the expansion to be valid is $|\xi\alpha| \ll 1$, where $\overline{N}_{in} = |\alpha|^2$ is the incident average photon number per mode, chosen accordingly to always fulfill the condition. All subsequent numerical analysis are done using $\overline{N}_{in} \leqslant 10^{11}$ and a fifth-order expansion of the operators Eq. (8). The associated errors on the operators is about $6.4\,\%$ and even smaller on the estimation of the gains, variances and $S$.

When only one of the three modes is excited by a bright coherent field $|\alpha\rangle$ containing $\overline{N}_{in}$ photons in average, no three-body quantum entanglement was found for any of the analyzed cases. If for example mode 3 is seeded, we can replace the operator $\hat{a}_3$ by its classical amplitude field counterpart $\imath E_3$, chosen to be complex for convenience, assuming that $E_3$ is a real number. The Hamiltonian becomes $\hat{H} = \imath \hbar \kappa (\hat{a}_1 \hat{a}_2 - \hat{a}_1^\dagger \hat{a}_2^\dagger)$, where $\kappa = gE_3$. This Hamiltonian describes SPDC, where a pump photon is converted into twin-photons. In the third-order configuration, however, the down-conversion process is proportional to an effective second-order susceptibility proportional to the product $\chi^{(3)} E_3$, meaning that the efficiency depends on the seeding level.

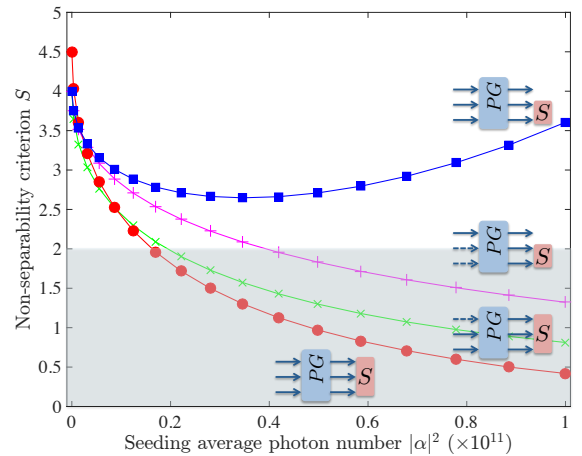

FIG. 1. Evolution of the non-separability criterion $S$ as a function of the seeding average photon number for $|\xi| = 1.75 \times 10^{-6}$ and for different injection cases. The gray area highlights the quantum entanglement region. The Triple-Photon Generation (TPG) boxes show the different schemes with the seeded (full arrows) and non-seeded (dashed arrows) modes. The $S$-boxes indicate the modes concerned by the $S$ measurement. Magenta line with (+) markers: single seeding case. Green line with (×) markers: double seeding case. Red line with (●) markers for the fully seeded at $\theta_1 = 0$, $\theta_2 = \theta_3 = \pi$, and for $\beta = \sqrt{2}$. The phase of the seeding coherent state is $\phi = \pi/2$. Blue line with (■) markers: same as the red curve but for quadratures $\hat{u} = \hat{Q}_1 + \hat{Q}_2$ and $\hat{v} = \hat{P}_1 - \hat{P}_2$.

Figure 1 allows to analyze the bipartite non-separability criterion $S$ for these twins. The magenta continuous line with the (+) markers shows $S$ for modes 1 and 2, using the quadratures $\hat{u} = \hat{Q}_1 + \hat{Q}_2$ and $\hat{v} = \hat{P}_1 - \hat{P}_2$, as a function of the seeding average photon number. The quadratures are determined from the exact solutions of Eq. (2). The criterion $S$ starts at 4, which is the classical limit, i.e. the sum of the four quadrature variances, then decreases as the seeding gets stronger. The gray area beneath $S = 2$ indicates the region in which quantum entanglement is present, meaning that the measurement of the quadrature $\hat{Q}_2$ (respectively $\hat{P}_2$) allows to know $\hat{Q}_1$ (respectively $\hat{P}_1$) better than the standard quantum limit associated with the shot noise of a coherent state [23]. Figure 1, shows that $S$ gets into the quantum entanglement region. The correlations get stronger as the seeding increases. In a sense, the single mode seeded TPG is a re-configurable two-body entanglement source where the seeded field acts as a control parameter: it changes the strength of the interaction and selects which quadratures are entangled depending on the seeding phase. When $E_3 \to 0$, it reaches the fluorescence case, which can be seen as an effective second-order process driven by the quantum fluctuations of mode 3. Actually this reasoning helps us to understand the origin of the missing three-body quantum entanglement. In-deed, each eigenvalue resulting from the measurement of the seeding mode, even at shot noise, will reset the twin-photon generation process.

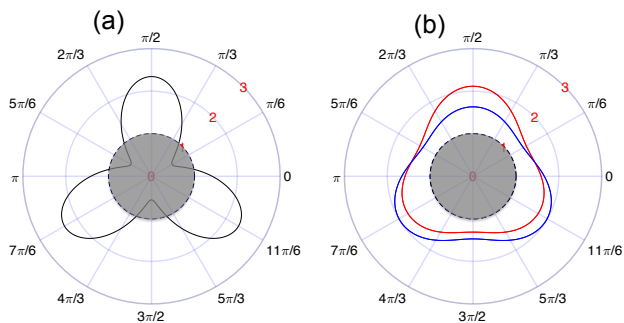

FIG. 2. Gain (a) and $\hat{P}_k$ (red) and $\hat{Q}_k$ (blue) variances (b) calculated using the fifth-order expansion of the operators as a function of the phase of the seeding fields, for $|\xi| = 1.75 \times 10^{-6}$ and $\overline{N}_{in} = 4 \times 10^{10}$. The gray circles indicate the unity gain in (a) and the standard quantum limit in (b).

The case of two modes coherently seeded TPG is analyzed hereafter. This situation corresponds to the experimental arrangement described in [21]. The input state of the system is $|\psi\rangle = |0_1, \alpha_2, \alpha_3\rangle$. The most trivial linear combinations in Eq. (3) to be considered is for $h_1 = g_1 = 1$, $h_2 = h_3 = 1/\sqrt{2}$ and $g_2 = g_3 = -1/\sqrt{2}$. All the local oscillator phases are $\theta_k = 0$. Any other phase does not influence the final result except by shifting the minimum value of $S$ in the phase space. The chosen parameters suggest that the quantum properties of the non-injected mode 1 are investigated through the measurements of both seeded modes 2 and 3. The green line with ($\times$) markers in Fig. 1 represents the corresponding $S$. Starting at 4, it decreases as the seeding grows. At $\overline{N}_{in} \simeq 2 \times 10^{10}$, $S$ goes below $f_p = 2$ indicating genuine three-body quantum entanglement. In this regime, measuring both modes 2 and 3 is necessary to know mode 1 better than the standard quantum limit.

*Fully seeded triple-photon generation.* This section focuses on the case where the three modes are initially excited. The input state of the system is $|\psi\rangle = |\alpha_1, \alpha_2, \alpha_3\rangle$. As for the fully seeded second-order parametric interaction, it is necessary to consider the phase of the different modes at the input of the nonlinear material relative to the pump phase. We also consider identical coherent states seeding the three modes, so that $\alpha_i = |\alpha|e^{i\phi}$ for each mode.

We start by looking at the evolution of the gain $G = \overline{N}_{out}/\overline{N}_{in}$, where $\overline{N}_{out} = \langle \hat{A}_k^\dagger \hat{A}_k \rangle$ and $\overline{N}_{in} = \langle \hat{a}_k^\dagger \hat{a}_k \rangle$ are the average photon number in each mode at the output and the input of the nonlinear medium. We also calculate the variance of the quadratures $\hat{P}_k$ and $\hat{Q}_k$ of each mode. Figure 2.(a) represents the gain $G$ for $\overline{N}_{in} = 4 \times 10^{10}$ in polar coordinates as a function of the phase $\phi$ of the seeding. It shows amplification ($G > 1$) and deamplification ($G < 1$) regimes and features three lobs subsequent to the $e^{i3\phi}$ dependance of $\overline{N}_{out}$. The maxima are located at $\phi = \pi/2$, $7\pi/6$ and $11\pi/6$ corresponding to the phases for which the down-conversion TPG process $\omega_p \to \omega_1 + \omega_2 + \omega_3$ is enhanced and achieving thus gains greater than unity. The minima are obtained for $\phi = \pi/6$, $5\pi/6$ and $3\pi/2$. They correspond to gains smaller than unity. Here, it is the sum frequency process $\omega_1 + \omega_2 + \omega_3 \to \omega_p$ which predominates. Figure 2.(b) shows the variances of the quadratures $\hat{P}_k$ (red) and $\hat{Q}_k$ (blue) in the same conditions of Fig. 2.(a). The most interesting feature, besides the presence of the three lobes, is that the variances are always above the shot noise (gray circle) when the modes are measured separately. This behavior is equivalent to non-degenerate twin-photon generation where signal and idler modes exhibit superpoissonian fluctuations.

Looking for tripartite quantum entanglement, we consider the following particular linear combinations of the operators of Eq. (3).

$$\begin{aligned}\hat{u} &= \hat{Q}_1 + \frac{1}{\beta\sqrt{2}}(\hat{Q}_2 + \hat{Q}_3) \\ \hat{v} &= \hat{P}_1 - \beta\frac{1}{\sqrt{2}}(\hat{P}_2 + \hat{P}_3),\end{aligned} \quad (11)$$

where we have introduced an extra free parameter $\beta$ for helpful optimization. Our analysis shows that for each phase $\theta_1$ of the local oscillator used for detecting mode 1, it exists a couple of phases $\theta_2$ and $\theta_3$ for detecting modes 2 and 3 that minimizes $S$, reaching values below the non-separability limit $f_p = 2$. We also show that further optimisation can be obtained through the choice of the adjustable parameter $\beta$. The red line with ($\bullet$) markers in Fig. 1 shows the evolution of $S$ as a function of the seeding average photon number per mode. It is obtained in the amplification regime. The local oscillator phases are $\theta_1 = 0$ and $\theta_2 = \theta_3 = \pi$. As the seeding increases, the red curve goes into the grey region below the limit $f_p = 2$, demonstrating genuine three-body quantum entanglement. This means that a measurement of both modes 2 and 3 is necessary to gain information on mode 1 better than the shot noise limit. Indeed, if one tries to determine mode 1 using only mode 2 for example by measuring the variances of the quadratures $\hat{u} = \hat{Q}_1 + \hat{Q}_2$ and $\hat{v} = \hat{P}_1 - \hat{P}_2$, then $S > 2$, as shown by the blue line with ($\blacksquare$) markers in Fig. 1.

*Conclusion.* In our analysis of parametric TPG, some counterintuitive results are found when comparing with the well-known twin-photon second-order equivalent. The first counterintuitive result is obtained for the pure spontaneous case for which we demonstrate in all cases the absence of three-body quantum entanglement. The more important finding is for the fully seeded TPG. It exhibits a phase-dependent gain, behaving similarly to the well-known second-order phase-sensitive parametric amplifier. But it also appears as an efficient way to generate genuine TPS quantum entanglement. Finally, an additional property corresponds to the increase in efficiency with the amplitude of the seeding classical field. All the parameters considered in these calculations are compatible with existing experimental results at the classical level for bulk materials. It is obvious that the overall efficiency could be greatly improved in the guided and integrated configurations.

This three-body genuine quantum entanglement constitutes a real asset for a new generation quantum cryptography and other quantum information protocols. We also stress that in the quest for the realization of continuous variable GHZ states, genuine entanglement of pure third-order TPS can offer the missing ingredient.

* kamel.bencheikh@c2n.upsaclay.fr
[1] D. M. Greenberger, M. A. Horne, A. Shimony, and A. Zeilinger, Am. J. Phys. **58**, 1131 (1990).
[2] W. Dür, G. Vidal, and J. I. Cirac, Phys. Rev. A **62**, 062314 (2000).
[3] S. Cheng and M. J. W. Hall, Phys. Rev. Lett. **118**, 010401 (2017).
[4] D. Bouwmeester, J.-W. Pan, M. Daniell, H. Weinfurter, and A. Zeilinger, Phys. Rev. Lett. **82**, 1345 (1999).
[5] A. S. Coelho, F. A. S. Barbosa, K. N. Cassemiro, A. S. Villar, M. Martinelli, and P. Nussenzveig, Science **326**, 823 (2009).
[6] H. Hübel, D. R. Hamel, A. Fedrizzi, S. Ramelow, K. J. Resch, and T. Jennewein, Nature **466**, 601 (2010).
[7] L. K. Shalm, D. R. Hamel, Z. Yan, C. Simon, K. J. Resch, and T. Jennewein, Nature Phys. **9**, 19 (2013).
[8] D. R. Hamel, L. K. Shalm, H. Hübel, A. J. Miller, F. Marsili, V. B. Verma, R. P. Mirin, S. W. Nam, K. J. Resch, and T. Jennewein, Nature Photon. **8**, 801 (2014).
[9] J. Douady and B. Boulanger, Opt. Lett. **29**, 2794 (2004).
[10] K. Bencheikh, F. Gravier, J. Douady, J. A. Levenson, and B. Boulanger, C. R. Physique **8**, 206 (2007).
[11] A. Efimov, A. J. Taylor, F. G. Omenetto, J. C. Knight, W. J. Wadsworth, and P. S. J. Russell, Opt. Express **11**, 2567 (2003).
[12] K. Bencheikh, S. Richard, G. Mélin, G. Krabshuis, F. Gooijer, and J. A. Levenson, Opt. Lett. **37**, 289 (2012).
[13] A. Cavanna, F. Just, X. Jiang, G. Leuchs, M. V. Chekhova, P. S. Russell, and N. Y. Joly, Optica **3**, 952 (2016).
[14] S. C. Warren-Smith, J. Wie, M. Chemnitz, R. Kostecki, H. Ebendorff-Heidepriem, T. M. Monro, and M. A. Schmidt, Opt. Express **24**, 17860 (2016).
[15] S. Richard, K. Bencheikh, B. Boulanger, and J. A. Levenson, Opt. Lett. **36**, 3000 (2011).
[16] M. Corona, K. Garay-Palmett, and A. B. U'ren, Opt. lett. **36**, 190 (2011).
[17] M. Corona, K. Garay-Palmett, and A. B. U'Ren, Phys. Rev. A **84**, 033823 (2011).
[18] S. L. Braunstein and A. K. Pati, Quantum information with continuous variables (Springer Science & Business Media, 2012).
[19] A. Dot, A. Borne, B. Boulanger, K. Bencheikh, and J. A. Levenson, Phys. Rev. A **85**, 023809 (2012).
[20] P. van Loock and A. Furusawa, Phys. Rev. A **67**, 052315 (2003).
[21] F. Gravier and B. Boulanger, J. Opt. Soc. Am. B **25**, 98 (2008).
[22] See supplemental material at [URL] for details on the calculation of the non-separability criterion $S$ in the case of spontaneous triple-photon generation and details on the validity of the operators expansion.
[23] Z. Y. Ou, S. F. Pereira, H. J. Kimble, and K. C. Peng, Phys. Rev. Lett. **68**, 3663 (1992).